\documentclass[10pt,preprint,a4paper]{aastex}

\usepackage{amsmath}                
\usepackage{amsfonts}               
\usepackage{amssymb}                
\usepackage{epsfig}                 

\newcommand{\s}{{~\rm s}}
\newcommand{\km}{{~\rm km}}

\newcommand{\erg}{{~\rm erg}}

\newcommand{\days}{{~\rm day}}

\begin{document}

\title{A BINARY SCENARIO FOR THE PRE-EXPLOSION OUTBURST OF THE SUPERNOVA 2010mc}

\author{Noam Soker\altaffilmark{1}}

\altaffiltext{1}{Department of Physics, Technion -- Israel Institute of Technology, Haifa 32000, Israel; soker@physics.technion.ac.il}

\begin{abstract}
I raise the possibility that the pre-explosion outburst (PEO) of the type IIn supernova 2010mc (PTF~10tel)
was energized by mass accretion onto an O main-sequence stellar companion.
According to this suggestion the SN progenitor suffered a rapid expansion within months before explosion.
The expansion was driven by leakage of energy from the core where vigorous oxygen nuclear burning takes place within a year prior to explosion.
This expansion triggered mass transfer onto the secondary star.
Most of the extra energy of the outburst comes from the accretion of $M_{\rm acc} \simeq 0.1 M_\odot$ onto the secondary star.
As well, the gas outflowing at $v_{\rm ej} \sim 2000 \km \s^{-1}$ was launched from the accreting secondary star, most likely
in a bipolar outflow.
The binary model can account for the slower circumstellar medium that was ejected at earlier times, and explain the red-shifted peak of the
H$\alpha$ emission at 5.8 days past explosion.
I compare some properties of the PEO of SN~2010mc to those of other stellar eruptions, such as the stellar merger event V838~Monocerotis
and the nineteenth century Great Eruption of the massive stellar binary system $\eta$ Carinae.
I speculate that all Type IIn supernovae owe their dense circumstellar gas to binary interaction.
\end{abstract}


\section{INTRODUCTION}
\label{sec:introduction}

With better sky coverage in recent years an accumulating number of core collapse supernovae (CCSNe) are observed to have pre-eruptions (outbursts)
from few years to just a month before the explosion.
The recent example is SN~2010mc (also known as PTF~10tel) who had a pre-explosion outburst (PEO) about a month prior to explosion \citep{Ofeketal2013b}.
SN~2010mc itself is a Type IIn supernova (SN) where the ejecta is thought to interact with close circumstellar matter (CSM).
In the case of SN~2010mc the early post-explosion interaction is with gas that was ejected in the PEO, but
slower gas expelled during earlier mass loss episodes resides further out \citep{Ofeketal2013b}.

Another interesting and enigmatic object is SN~2009ip, that had a major outburst in September 2012 (outburst 2012b; \citealt{Mauerhan2012}),
and pre-eruptions in 2009, 2011, and August 2012 (outburst 2012a;
e.g., \citealt{Berger2009, Maza2009, Drake2012, Mauerhan2012, Pastorello2012, Levesque2012}).
\cite{Mauerhan2012} attributed the 2012 outbursts to a CCSN explosion. \cite{Pastorello2012} put to question this suggestion,
and \cite{SokerKashi2013} further argued that the 2012b was not a CCSN but rather a mergerburst event where two massive stars merged.
The cause of the major 2012b outburst of SN~2009ip is still an open question, but the pre-outbursts are similar in duration and energy
to the pre-outburst of SN~2010mc.

In some cases the pre-eruptions are compared to the nineteenth century Great Eruption (GE) of $\eta$ Carinae
(e.g., \citealt{Foley2011} and \citealt{Smithetal2010} for SN~2009ip).
As the sharp peaks in the GE of $\eta$~Car were triggered by binary interaction at periastron passages
\citep{Damineli1996, KashiSoker2010a, SmithFrew2011}, the GE is related to a binary interaction.
The extra energy of the GE of $\eta$ Car might have been originated from accretion of mass onto the secondary O star
\citep{KashiSoker2010a}.

I note the similarity in the energy and time scales of the one-month PEO of SN~2010mc to the sharp peaks of the GE of $\eta$ Car
and to other Intermediate Luminosity Optical Transients (ILOTs; also termed intermediate luminosity red transients and luminous red novae),
such as 2008S (see properties of these in \citealt{SmithFrew2011}).
Motivated by these similarities I propose in section \ref{sec:binary} that the PEO of SN~2010mc was powered
by mass accretion onto an O main sequence (MS) star.
My proposed scenario is different than the one suggested by \cite{Pastorelloetal2007} to account for the two-years PEO of
SN~2006jc. In their model a WR star exploded as the SN, while a massive luminous blue variable (LBV) companion was responsible to the PEO.
In the model proposed here the companion is a MS star that accreted mass from the SN progenitor.

\cite{Ofeketal2013b} argue that the PEO properties are consistent with being powered by
waves that are excited by vigorous core-convection during the oxygen and/or neon burning phases in the months
prior to explosion, as suggested by \cite{QuataertShiode2012}.
In section \ref{sec:expan} I suggest that the main effect of such waves might be huge envelope expansion rather than mass expulsion.
A short summary is in section \ref{sec:summary}.

\section{THE BINARY MODEL}
\label{sec:binary}

I start by mentioning six objects: the GE of $\eta$ Car, the SN-impostor SN~2008S, the 2011 and 2012a outbursts of SN~2009ip prior to the
major 2012b outburst, the mergerburst event V838 Mon, the outburst of NGC~300~OT2008-1, and the pre-explosion outburst (PEO) of SN~201mc.
All these are grouped into the optical-time stripe (OTS), a stripe occupying the region between novae and SNe
in a energy-time diagram. Most are collectively referred to as intermediate luminosity optical transients (ILOTs),
while several objects are LBV major eruptions. There is an overlap between the LBV and ILOTs on the OTS \citep{Kashisoker2010b}.
The PEO of SN~2010mc, with a duration of $\tau \sim 1$~month and a total energy of $E_{\rm tot} \simeq 10^{48} \erg$
\citep{Ofeketal2013b}, is located on the energy-time diagram close to V838~Mon and SN~2008S \citep{Kashi2013}.

The light curves of some of these were compared to each other. Examples include the comparison of
the entire GE of $\eta$ Car with the light curves of V838~Mon and NGC~300~OT \citep{Kashietal2010},
the sharp peaks of the GE of $\eta$ Car with the light curves of NGC~300~OT and of SN~2008S \citep{SmithFrew2011},
the 2009 outburst of SN~209ip with the light curves of $\eta$ Car, NGC~300~OT, and SN~2008S \citep{Smithetal2010},
and the light curves of the 2012a+2012b outbursts of SN~2009ip with that of V838 Mon \citep{SokerKashi2013}.
\cite{Kashisoker2010b} suggest that all these outbursts are powered by gravitational energy released in binary systems.
In some ILOTs a merger occurs and one of the stars does not survive the transient event, e.g., V838~Mon \citep{TylendaSoker2006}
and V1309~Scorpii \citep{Tylendetal2011a}.
In some other ILOTs a rapid and short mass transfer process takes place and the two stars survive the transient event, e.g., the GE $\eta$ Car.

The PEO energy of SN~2010mc, $\sim  10^{48} \erg$, can be released by a mass of $M_{\rm acc} \simeq 0.1 M_\odot$ that is accreted onto a secondary
MS star of $(M_2,R_2)=(20 M_\odot, 6 R_\odot)$, where the secondary properties are as \cite{SokerKashi2013} proposed for the secondary of SN~2009ip.
The rapid mass transfer process from the primary to the secondary is triggered by the primary's envelope expansion.
The reason that the PEOs occur within months (in the case of SN~2010mc it is one month) before explosion is that the large expansion is caused by the powerful oxygen
burning that occurs within about a year before explosion. A small fraction of the released nuclear energy is assumed to leak to the envelope
and cause its expansion (see section \ref{sec:expan}).
The accreted mass is similar to the estimated mass that was ejected in the month-long PEO \citep{Ofeketal2013b}.
The escape velocity from such a secondary star is $1100 \km \s^{-1}$, and such a star launching a bipolar outflow can account for
the observed velocities of $\sim 2000 \km \s^{-1}$.
For example, a small fraction of the mass ejected in the bipolar outflow of the GE of $\eta$ Car is moving at high velocities \citep{Weisetal2001}
of up to $\sim 5000 \km \s^{-1}$ \citep{Smith2008}.

Many ILOTs, including the PEO of SN~2010mc and the 2012a outburst of SN~2009ip,
are located below SNe in the energy-time diagram \citep{Kashi2013}. Namely, they have a similar duration of several weeks to few months,
but the total energy (kinetic + radiation) involved is $\sim 3-4$ orders of magnitudes lower than in SNe.
In many of these ILOTs the ejected mass is $\sim 0.01- {\rm few} \times 0.1 M_\odot$, e.g., the PEO of SN~2010mc \citep{Ofeketal2013b},
the 2012a outburst of SN~2009ip \citep{Ofeketal2013a}, and V838-Mon \citep{TylendaSoker2006}.
The ejection velocities in these ILOTs is $\sim 10^3 \km \s^{-1}$ which in the binary model is determined by the escape velocity from
OB MS stars.
The width of a supernova light curve (ignoring interaction with a CSM) is approximately given by \citep{Arnett1982, Mazzalietal2001}
\begin{equation}
    \tau_{\rm w} \sim \sqrt{\frac {\kappa  M_{\rm ej}}{10 v_{\rm ej} c} }
    \simeq 20 \left( \frac {\kappa}{0.34} \right)^{1/2}
    \left( \frac {M_{\rm ej}}{1 M_\odot} \right)^{1/2}
    \left( \frac {v_{\rm ej}}{10^4 \km \s^{-1} } \right)^{-1/2} ~{\rm days}
\label{eq:tauw1}
\end{equation}
where $\kappa$ is the opacity, $M_{\rm ej}$ and $v_{\rm ej}$ are the ejected mass and its velocity, respectively,  and $c$ is the light speed.
The mass ejected in ILOTs is $\sim 0.01-0.1$ times that in SNe, and the ejection velocity is $\sim 0.05-0.2$ times that in SNe. From equation (\ref{eq:tauw1})
we find the duration of ILOTs with $M_{\rm ej} \simeq 0.01- {\rm few} \times 0.1 M_\odot$ and $v_{\rm ej} \simeq 10^3 \km \s^{-1}$ to be similar
to the width of SN light curves, if the outburst width is determined by photon diffusion.
This is not always the case as in many cases the engine of an ILOT continues to work for a longer time, as in the GE of $\eta$ Car.
Non the less, equation (\ref{eq:tauw1}) might motivate further research in that direction.

One of the outcomes of the binary model is a highly non-spherical mass loss ejection, e.g., the formation of bipolar lobes as in
$\eta$ Car, or the formation of a dense equatorial ring (termed the equatorial skirt in $\eta$ Car).
There might be an indication for a non-spherical mass ejection in SN~201mc.
At $t=5.8 \days$ the H$\alpha$ peak is somewhat red-shifted relative to the peaks at later times (it seems to be a real effect; Ofek, E. O. 2013, private communication).
This can be understood if the closer side that is responsible for the blue-shifted emission, is partially obscured.
This is hard to account for in a spherical geometry, where the peak is expected in the blue-shifted side, but naturally appears
in bipolar lobes or equatorial mass concentration (a torus or a ring) in the following way.
At $t= 5.8 \days$ the slow moving gas did not reach its peak emission yet (as evident from Fig. 2 of \citealt{Ofeketal2013b}),
implying that it is partially ionized and/or it is still obscured by dust.
The ionizing radiation from the explosion destroys first the dust closer to the star. In a torus (ring) inclined to the observer the
side further away will be less obscured by dust, hence red-shifted emission will be stronger.

\section{PRIMARY EXPANSION}
\label{sec:expan}

Some ILOTs are triggered by the evolution of the binary orbit, e.g., the shrinkage of the orbit in V1309~Scorpii \citep{Tylendetal2011a}.
Others ILOTs  are triggered by some instability of the primary star, as the suggestion for the triggering of LBV major eruptions,
such as the GE of $\eta$ Car \citep{KashiSoker2010a} and the seventeenth century eruption of P~Cygni \citep{Kashi2010}.
In some cases the two mechanisms might be connected, e.g., in the binary model for SN~2009ip \citep{SokerKashi2013}
the orbit was shrinking due to tidal interaction with the primary.

In the proposed binary model for SN~2010mc the PEO was triggered by instability in the primary star.  Such an instability could
have been driven by waves excited by core convection as proposed by \cite{QuataertShiode2012}.
These waves carry a small fraction of the nuclear energy released in the vigourous burning of oxygen.
According to \cite{QuataertShiode2012} the main effect of the waves is to substantially enhance the mass loss rate.
Here I suggest that the main effect of such waves might be an envelope expansion due to
energy deposition in inner envelope layers.
Envelope expansion due to energy leakage from powerful nuclear burning in the core has been suggested to occur during
core-helium flash of low mass stars (e.g., \citealt{Bearetal2011}).
It is such an envelope expansion that is at the heart of the process discussed here.

\cite{Bearetal2011} injected energy at the base of the convective envelope of a $0.9 M_\odot$
red giant branch (RGB) star at a rate of $10^5 L_\odot$ for seven years.
The total energy injected at the bottom of the convective region, just above the hydrogen burning shell, was $8.5 \times 10^{46} \erg$.
\cite{Bearetal2011} found that within these seven years the photosphere grew from
$100 R_\odot$ to $\sim 700 R_\odot$, and the outer convective region grew from
$100 R_\odot$ to $\sim 400 R_\odot$.
Namely, the convective envelope in their RGB model could absorb $\sim 10^{47} \erg$ by expanding and without being disrupted.
Such a huge expansion will of course be accompanied by mass loss. But the point I raise here that the main effect of a large
energy deposition to the envelope over a time shorter than the thermal time scale might be envelope expansion.

Let me examine the implications of the wave propagation in the envelope.
The waves carry energy with a luminosity of $L_{\rm wave} \simeq 4 \pi r^2 e_w c_s$, where $e_w$ is the
energy density of the waves and $c_s$ is the sound speed.
The local thermal energy density of the gas is $e_{\rm th} \simeq \rho c_S^2$, such that
\begin{equation}
    \frac {e_w}{e_{\rm th}} \simeq
    \frac{L_{\rm wave}}{L_{\rm max,conv}},
\label{eq:enden1}
\end{equation}
where $L_{\rm max,conv} = 4 \pi \rho r^2 c_s^3$ is the maximum energy that subsonic convection can carry as used by \cite{QuataertShiode2012}.
\cite{QuataertShiode2012} give the values of $L_{\rm wave}$ and $L_{\rm max,conv}$ for their model, from which I deduce that
${e_w}/{e_{\rm th}}=0.1$,  $0.3$, and $1$ at $r/R_\ast =0.07$, $0.1$, and $0.3$, respectively,
where $R_\ast=1700 R_\odot$ is the stellar radius of their model.
This implies that inner layers of the envelope at $r \simeq 0.1 R_\ast$ are substantially influenced by the waves, and will expand and absorb energy.
As the expansion time scale is similar to the waves-propagation time, I suggest that the envelope has time to respond and absorb most of the
energy carried by the waves, and that strong shocks will not be formed by the waves.
Definitely the influence of waves on the envelope requires further much detail studies.

Once the primary envelope expands, the binary companion can find itself inside the bloated atmosphere, or very close to it, and a rapid mass transfer process
can take place. In this model the secondary is already close to the primary.
To prevent the secondary from entering the primary envelope at earlier stages, the secondary must be a massive star,
$M_2 \ga 0.1 M_1$. The secondary is a MS star of type B or O in the model proposed here.
The accreted mass has enough angular momentum to form and accretion disk, and to launch jets, as in the binary model for the formation of
the the Homunculus$-$the bipolar nebula of $\eta$ Car \citep{Soker2001}.

\section{DISCUSSION AND SUMMARY}
\label{sec:summary}

There are two basic schools of thought when it comes to explain the heterogenous group of erupting objects having luminosity between
novae and supernovae (SNe). This group is composed mainly of ILOTs (for intermediate luminosity optical transients; also termed intermediate
luminosity red transients or red luminous novae) and major outbursts of luminous blue variables (LBV).
In the energy-time diagram (ETD\footnote{For an updated ETD see http://physics.technion.ac.il/$\sim$ILOT/}) they are located on a strip, the optical transient stripe \citep{Kashi2013}.
One school attributes the eruptions to effects solely due to single star evolution, e.g., the porous-atmosphere
model that was suggested by \cite{Shaviv2000} for the super-Eddington nineteenth century Great Eruption (GE) of $\eta$ Car,
and the single star model of SN~2008S and NGC~300~OT \citep{Kochanek2011}.
The other school argues that binary interaction is at the heart of these eruptions, and the source of energy is gravitational energy released by
either mass transfer or stellar merger. Examples include mass transfer processes as in the binary model for the GE of $\eta$ Car
\citep{Soker2001, KashiSoker2010a}, in NGC~300~OT \citep{Kashietal2010}, and in the early outbursts of SN~2009ip \citep{SokerKashi2013}.
A binary merger model (mergerburst) was proposed for the final outburst of SN~2009ip \citep{SokerKashi2013}.

The new report by \cite{Ofeketal2013b} on the pre-explosion outburst (PEO) of SN~2010mc adds another `battle ground'' for the two schools
to apply their processes.
\cite{Ofeketal2013b} adopt the porous-atmosphere model \citep{Shaviv2000, Shaviv2001} in a single-star scenario to account for the PEO of SN~2010mc.
In this paper I suggest (section \ref{sec:binary}) instead that the main energy source was mass accretion onto an O main sequence
(or a slightly evolved) companion.
The mass transfer was triggered by the expansion of the primary star (section \ref{sec:expan}).
Such an expansion could have been driven by waves excited by the oxygen burning in the core \citep{QuataertShiode2012}.
The binary model can account for the slower circumstellar medium that was ejected at earlier times, and explain the red-shifted peak of the
H$\alpha$ emission at $t=5.8 \days$ reported by \cite{Ofeketal2013b}.

By studying groups of CCSNe \cite{Arcavi2012} concluded that binary interaction plays a major role in the
mass loss process from progenitors of Type Ib CCSNe. In these cases binary companions that enter a common envelope phase might
account for the striping of the hydrogen-rich envelope.
Here I argue that binary companions that avoid a common envelope phase, at least to the very end, are behind the PEOs.

Such main-sequence binary companions might account for ILOTs of low mass stars that ends their nuclear-active lives in a planetary nebula.
\cite{SokerKashi2012} suggested that some asymptotic giant branch stars can have an ILOT event energized by accretion onto a companion.
The accretion disk that is formed around the companion launches two jets that shapes the descendant planetary nebula to a bipolar one.
\cite{SokerKashi2012} suggested this scenario to account for the properties of the bipolar  planetary nebula NGC~6302, that has some similarities with the
ILOT NGC~300~OT.
\cite{BoumisMeaburn2013} suggested that the bipolar planetary nebula KjPn~8 was also shaped by an ILOT event.

More observations, searching for binary systems and establishing the circumstellar medium morphology, and more theoretical studies,
focusing on the evolution of massive stars and on mass transfer processes in massive binary stars, are required to tilt the balance in favor of one of
the two schools of thought.

I thank Amit Kashi for helpful suggestions.
This research was supported by the Asher Fund for Space Research at the Technion, and the US-Israel Binational Science Foundation.

{}

\end{document}